\documentclass[
 aip,
 amsmath,amssymb,
 reprint,%
]{revtex4-2}

\usepackage{graphicx}
\usepackage{dcolumn}
\usepackage{bm}

\usepackage[utf8]{inputenc}
\usepackage[T1]{fontenc}
\usepackage{mathptmx}
\usepackage{etoolbox}

\makeatletter
\def\@email#1#2{%
 \endgroup
 \patchcmd{\titleblock@produce}
  {\frontmatter@RRAPformat}
  {\frontmatter@RRAPformat{\produce@RRAP{*#1\href{mailto:#2}{#2}}}\frontmatter@RRAPformat}
  {}{}
}%
\makeatother
\begin{document}

\preprint{AIP/123-QED}

\title{Resource-efficient low-loss four-channel active demultiplexer for single photons}

\author{M. Dryazgov}
\affiliation{Quantum Technology Centre, Lomonosov Moscow State University, Moscow, Russia}

\author{Yu. Biriukov}
\affiliation{Quantum Technology Centre, Lomonosov Moscow State University, Moscow, Russia}

\author{I. Dyakonov}
\affiliation{Quantum Technology Centre, Lomonosov Moscow State University, Moscow, Russia}

\author{K. Taratorin}
\affiliation{Quantum Technology Centre, Lomonosov Moscow State University, Moscow, Russia}

\author{A. Korneev}
\affiliation{Quantum Technology Centre, Lomonosov Moscow State University, Moscow, Russia}

\author{M. Rakhlin}
\affiliation{Ioffe Institute, St. Petersburg, 194021, Russia}

\author{A. Galimov}
\affiliation{Ioffe Institute, St. Petersburg, 194021, Russia}

\author{G. Klimko}
\affiliation{Ioffe Institute, St. Petersburg, 194021, Russia}

\author{S. Sorokin}
\affiliation{Ioffe Institute, St. Petersburg, 194021, Russia}

\author{M. Kulagina}
\affiliation{Ioffe Institute, St. Petersburg, 194021, Russia}

\author{Yu. Zadiranov}
\affiliation{Ioffe Institute, St. Petersburg, 194021, Russia}

\author{A. Toropov}
\affiliation{Ioffe Institute, St. Petersburg, 194021, Russia}

\author{F. Bergmann}
\affiliation{Bergmann Messgeraete Entwicklung KG, Murnau, Germany}

\author{S. Straupe}
\affiliation{Quantum Technology Centre, Lomonosov Moscow State University, Moscow, Russia}%
\affiliation{Russian Quantum Center, Skolkovo, Moscow 143025, Russia}

\author{S. Kulik}
\affiliation{Quantum Technology Centre, Lomonosov Moscow State University, Moscow, Russia}
\affiliation{Laboratory of Quantum Engineering of Light, South Ural State University (SUSU), Russia, Chelyabinsk, 454080, Prospekt Lenina 76}

\date{\today}

\begin{abstract}
We report a design and implementation of a resource-efficient spatial demultiplexer which produces 4 indistinguishable photons with efficiency of 39.7\% per channel. Our scheme is based on a free-space storage/delay line which accumulates 4 photons and releases them by a controlled polarization rotation using a single Pockels cell. 
\end{abstract}

\maketitle


Linear-optical quantum computing requires an efficient method of on-demand generation of a definite number of indistinguishable photons which can be used to encode qubits, e.g. fed into a photonic circuit performing a quantum algorithm. The majority of the reported experimental implementations make use of either spontaneous parametric down-conversion (SPDC) in bulk crystals \cite{Zhong2018, Qian2023} or four-wave mixing in waveguides \cite{Qiang2018, Bao2023} in order to generate single photons. Recent advances in semiconductor quantum dot (QD) sources \cite{Senellart2016} and emergence of commercially available QD source chips (Quandela, Sparrow Quantum or AegiQ) propelled the spread of this type of sources in the field of optical quantum computing \cite{Istrati2020, Corrieli2022}. Fabrication of a set of identical QD sources is a notoriously hard task and hence the experimental challenge of using a single QD as a multiphoton source has emerged.

The SPDC process is easily adopted to generate several photons at once. A straightforward solution implies assembling $n$ identical SPDC photon pair sources to generate $2n$ photons with probability $p_{\mathrm{SPDC}}^{n}$. The drawback is an extremely low probability $p_{\mathrm{SPDC}}<<1$ of successful scattering. The $p_{\mathrm{SPDC}}$ can be substantially enhanced if the heralding feature of the SPDC photon pair source is used to implement the multiplexing principle. Detection of one photon of a pair heralds the successful generation of a pair and hence the second photon can be stored in an optical memory cell and released on demand. F.~Kaneda and P.~Kwiat provided the most prominent demonstration of this method~\cite{Kaneda2019}. Their implementation includes two Pockels cells (PC) one of which is used for storage and release of the heralded photon and the second one filters out the unheralded photons. As a result they show generation of pure and indistinguishable single photons with probability $p_{\mathrm{MUX}}=66.7(24)\%$ per cycle. The scheme can be cloned to produce several photons simultaneously at a cost of using many SPDC crystals, storage cells, single-photon detectors and fast switches (at least two Pockels cells per each multiplexed photon). The review \cite{MeyerScott2020} provides the comprehensive viewpoint on single-photon multiplexing techniques.

A state-of-the-art QD source is a single-photon emitter coupled to a microcavity. This concept makes close to deterministic generation of single-photons feasible in principle~\cite{Senellart2016, Tomm2021}. Multiphoton state preparation relies on a reciprocal idea of demultiplexing -- the stream of photons emitted by a dot is split into several independent spatial modes. Photons that were generated during $N$ consecutive clock cycles (or $N$ consecutive bursts of cycles) of the pump laser are forwarded towards delay lines which produce $N$ synchronized photon packs at the output~\cite{Hummel2019, Munzberg2022}. This scheme requires $N-1$ switches (Pockels cells, resonant phase modulators or other). The maximal amount of demultiplexed photons which has been demonstrated to date with this method is 20~\cite{Wang2019}.
 
An ideal demultiplexer should demonstrate low losses in switching and routing of photons, and maintain indistinguishability of photons after demultiplexing. It is highly desirable to design a resource-efficient demultiplexer circuit which includes the least possible number of switches, because each switch introduces loss mostly due to non-ideal switching extinction. The complexity of assembling, adjusting, and controlling the circuit grows significantly with the amount of switches and effectively increases the overall loss per demultiplexer channel. In~\cite{Lenzini2017} 4-channel demultiplexing was performed on a lithium niobate photonic chip. Although an on-chip demultiplexer demonstrates 78\% efficiency, the input and output coupling to a photonic chip reduces the total transmittance of the demultiplexer to 30\% only. This work demonstrates the prospect of developing fast on-chip modulators capable of individual pulse-picking which corresponds to the shortest possible delay lines. Further development of an integrated demultiplexer design led to a nanomechanical on-chip photon switch reported in \cite{Papon2019}. Despite being inferior to its electrooptic counterpart \cite{Lenzini2017} the nanomechanical switch demonstrated feasibility of packing both a source and a demultiplexer in a single technological platform without invoking the use of hybrid integration technologies.

\begin{figure*}[ht!]
	\includegraphics[width=2\columnwidth]{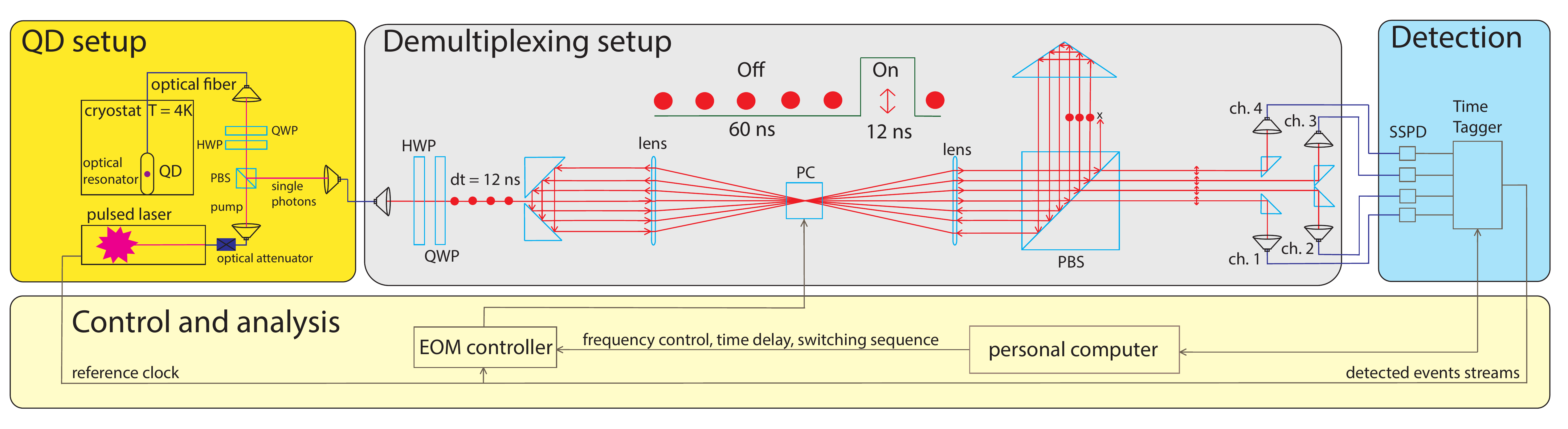}
	\caption{A schematic illustration of the experiment with the QD single photon source setup, the demultiplexing setup, the detection setup, and control and analysis setup. The demultiplexer optical circuit is feeded with a sequence of single photon pulses emitted by a QD in a microresonator source. The circuit stores four optical pulses and outputs them once the Pockels cell PC is turned on and the polarization of each pulse id rotated to the orthogonal state. The detailed description of the setup is given in Sec.~\ref{Experimental_method}.}
	\label{fig:setup}
\end{figure*}

The works \cite{Pan2017, Pan2019, Hummel2019} were first to demonstrate spatial demultiplexing of a single stream of photons produced by a single quantum dot. Both experiments employ the same layout of the demultiplexer -- the tree-like array of electrooptic switches (Pockels cells). This layout requires $N-1$ switches to separate an input stream into $N$ ports. The authors achieved high transmission per channel equal to approximately 83-85\%. The work~\cite{Pont2022} employs a fast acousto-optic deflector (AOD) to divert the input beam towards each output port. After that the photons could be spatially separated by first-order diffraction. For this purpose, the AOD was driven by three different RF frequencies within time windows $\Delta T$. Each frequency corresponds to deflection of the beam towards a designated output fiber coupler. An optical frequency shift induced by an AOD was much smaller than the bandwidth of the photons and thus did not affect their indistinguishability. The best reported transmission was 65\%. Here the time window $\Delta T = 320$~ns is defined by the speed of AOD switching. The longest delay line in this example equals to $960$~ns imposing a stringent requirement on the QD source itself: the photons separated by a $960$~ns interval should still maintain a high degree of indistinguishability. Another implementation was reported in \cite{Senellart2019, Keil2022}. The main difference is the use of resonantly-enhanced electrooptic modulators. In \cite{Senellart2019} the resonant frequency wasn't tuned to match the repetition rate of the pump laser and hence some of the photons were lost during the switching process. The work \cite{Keil2022} demonstrated a system where this problem is solved and the demultiplexer is capable of distributing each photon from the stream to a designated channel.

In this paper we present a resource-efficient four-beam demultiplexer requiring a single fast optical switch (Pockels cell (PC)) only. Although our demultiplexer does not beat the best reported probability to produce a photon at the output, it drastically reduces the amount of fast switches required for splitting the input photon sequence and eliminates the necessity to assemble long fiber delay lines in order to compensate for large delay times between the output photon bursts in each channel.   

\section{Experimental method} \label{Experimental_method}

We exploit a simple idea of storing and releasing a bunch of photons inside an optical loop. Our setup is shown in Fig.~\ref{fig:setup}. We use a semiconductor QD source in the micropillar configuration \cite{Rakhlin2022}. We pump the QD in a cross-polarized scheme using resonant picosecond pulses emitted at 82.6~MHz repetition rate at 918.83~nm wavelength. The more detailed description of the QD source is available in Appendix \ref{app:qd-source}. The single photons are coupled to a single-mode optical fibre and fed into the demultiplexer through the output coupler. The demultiplexer is an optical loop with a round-trip time equal to the time difference $\Delta t = 12.1$~ns between the consecutive pump pulses.  Each round-trip shifts the beam transversely in a horizontal direction by 3~mm on average (see Appendix \ref{app:demux-details}). A 1:1 telescope inside the loop serves two purposes: it minimizes the diffraction spread of the beam passing through the loop multiple times and focuses all the beams through the 3~mm aperture of the PC (Leysop). The currently used geometry of the optical components allows to store only 4 photons inside the loop. The half-wave and quarter-wave plates at the input adjust the polarization of the input photons to a horizontal state. After passing through the PC each photon reflects of the polarization beam splitter (PBS) and the right-angle prism and completes the round trip by passing the PC again. Each sixth clock cycle the PC rotates the polarization of the photons in the loop and hence releases an incoming photon and three previously stored ones. Ideally switching has to be done every fourth cycle because our circuit is designed to collect only four output photons, however hardware restrictions imposed by the high-voltage driver allow for a minimal interval between consecutive switching events of $70$~ns. The high-voltage driver (BME Bergmann) switches the cell using a 12~ns long voltage pulse. The driver is controlled by a homebuilt FPGA-based circuit which uses the pump laser as a source of the reference clock signal. The voltage pulse peak is synchronized to the moment when all four photons pass through the Pockels cell and the rise and fall fronts do not affect the neighbouring photons.

The photons with rotated polarization pass through the PBS and enter the circuit where each beam is separated from the other using knife-edge reflective prisms. After that photons are coupled to the single mode fibres. Each fiber coupler is mounted onto a translation stage in order to compensate for slight differences of the output channel lengths including the lengths of optical fibers. The output of each fiber coupler is sent either to superconducting single-photon detectors (SSPD) or to a balanced fiber beamsplitter which enables the observation of Hong-Ou-Mandel type quantum interference.

\section{Results}
\begin{figure}[tb]
	\includegraphics[width=\columnwidth]{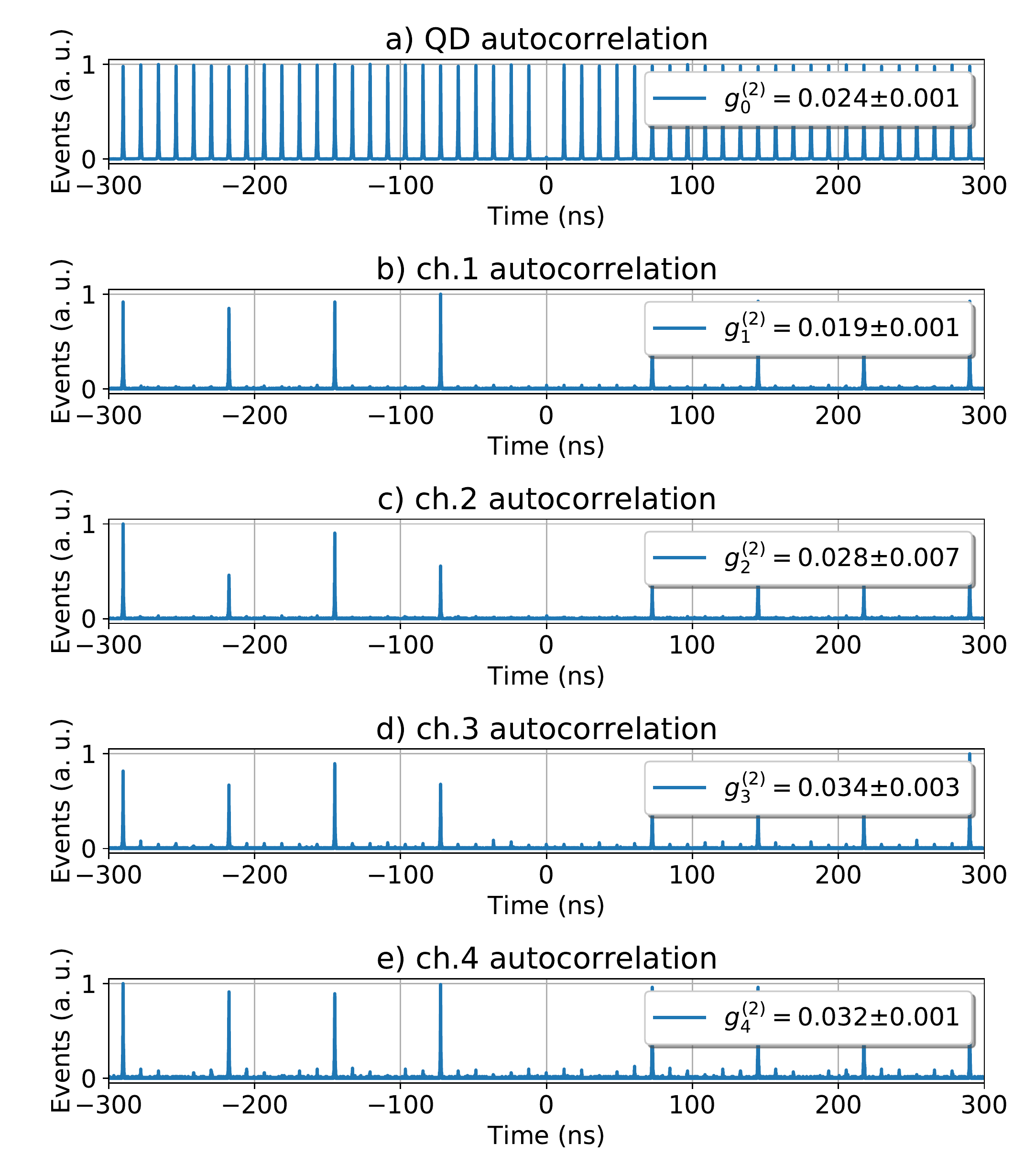}
	\caption{(a) Auto-correlation function measured for the single photons emitted by the QD. (b-e) Auto-correlation functions measured for the output signal at each of the 4 channels of the demultiplexer.}
	\label{fig:correlation}
\end{figure}

Firstly, we characterized the QD itself. Figure~\ref{fig:correlation}(a) presents the auto-correlation function $g^{(2)}(\tau)$ for our QD pumped in a resonant regime in a cross-polarized configuration. The observed single-photon purity is $g^{(2)}(0)=0.024 \pm 0.001$. The indistinguishability of the photons was asserted using a Hong-Ou-Mandel interferometer with $\Delta t = 12.1$~ns delay introduced into one of the arms. The average photon count rate detected using the SSPD was about 5~MHz.  

Then, we measured the auto-correlation functions for each output channel, the result is presented in Fig.~\ref{fig:correlation}(b--e). The auto-correlation histograms indicate the quality of splitting the input photon sequence. Small parasitic peaks can be witnessed between the demultiplexed photons. They originate from the imperfect polarization switching and finite PBS extinction ratio. Our configuration implies passing four beams through a single PC and each beam travels in a slightly different direction due to focusing. For this reason it is impossible to reach optimal performance of the PC for each beam. We decided to use the second output channel as a reference one because it was easier to find the best orientation of the PC. Thus the second channel auto-correlation function has minimal parasitic peaks.

\begin{figure}[tb]
	\includegraphics[width=\columnwidth]{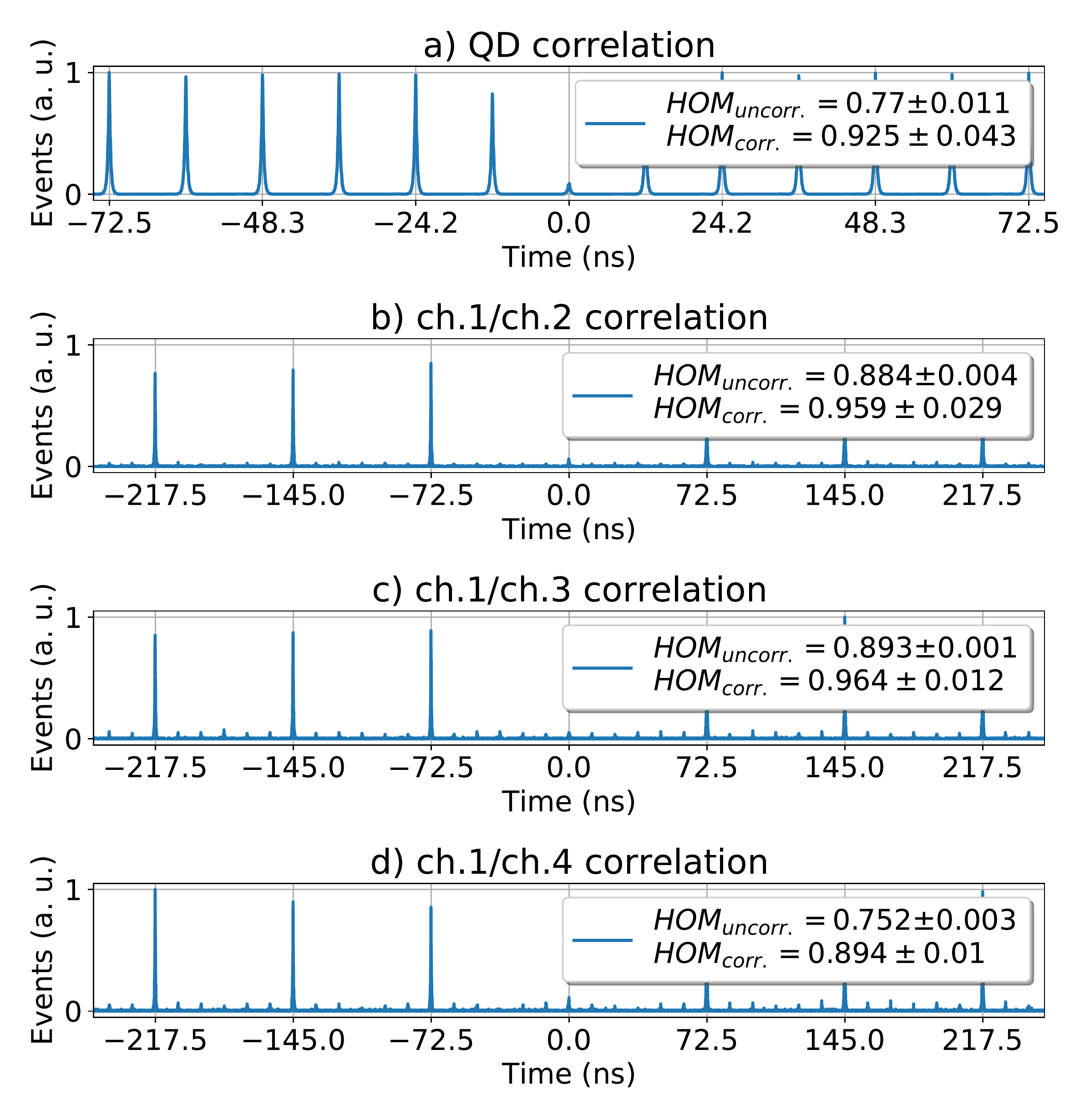}
	\caption{Uncorrected and corrected indistinguishability of the single photons from channels 1 and 2 (panel a) ), 1 and 3 (panel b) ) and 1 and 4 (panel c) ) of the demultiplexer. The indistinguishability value is correct in order to take account of non-ideal beamsplitter reflection and non-zero $g^{(0)}$ value of the interfering light. See Appendix \ref{app:auto-and-hom} for the detailed explanation.}
	\label{fig:HOM}
\end{figure}

Next we tested the indistinguishability of the photons after they have been split into different channels by measuring Hong-Ou-Mandel (HOM) interference visibility. We connected a balanced fiber beamsplitter to the outputs of a selected pair of channels and measured the second-order cross-correlation function $g^{(2)}(\tau)$. Since we used a single-mode fiber which doesn't preserve the polarization state, we had to place an additional pair of half- and quarter-wave plates to each channel and vary their orientation to compensate the unknown polarization rotation inside the fibers. The measurement results indicate that the indistinguishability of the photons remains on the same level as for the photons tested directly from the QD source. We measured raw indistinguishability values between channels 1 and 2 $HOM_{uncorr}^{12}=0.884$, channels 2 and 3 $HOM_{uncorr}^{23}=0.893$ and channels 3 and 4 $HOM_{uncorr}^{34}=0.752$. The corresponding values corrected for non-zero $g^{(2)}(0)$ and interferometer imperfections are $HOM_{uncorr}^{12}=0.982$, $HOM_{uncorr}^{23}=0.986$ and $HOM_{uncorr}^{34}=0.916$. We used a correction formula from \cite{Senellart2016} to infer the estimated value of the Hong-Ou-Mandel interference visibility (see Appendix \ref{app:auto-and-hom}). We note that raw HOM values $HOM_{uncorr}^{12}$ and $HOM_{uncorr}^{23}$ exceed the source indistinguishability possibly due to imbalanced detection efficiencies in the corresponding channels.

\begin{figure}[tb]
	\includegraphics[width=\columnwidth]{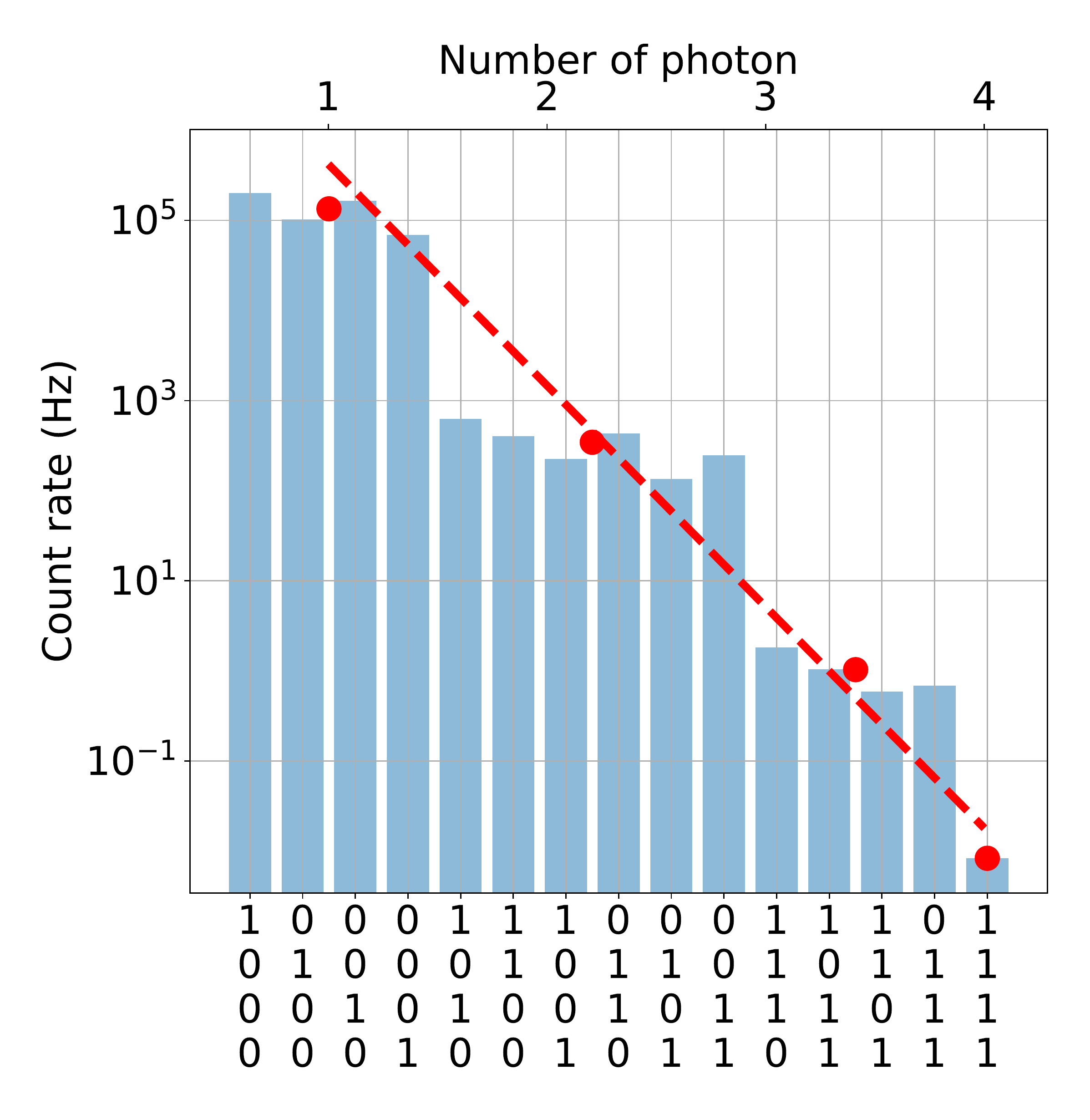}
	\caption{Count rates of different configurations of photon distributions among channels. The red dots are the average count rate values for the given number of photons. Red line is the approximation by an exponential function $p^n$}
	\label{fig:countrate}
\end{figure}

Finally, Fig.~\ref{fig:countrate} illustrates the detection rates of multiphoton events at the output of the demultiplexer. We used the data to estimate the efficiency across all output channels. The probabilities to detect $n$-photon events were fitted with an exponential function $p^{n}$, where $p$ is the probability to detect a photon at the output of the demultiplexer. The value $p$ is related to the source brightness $B=\nu/r$, where $\nu$ is the detected single-photon count rate at the output of the source and $r$ is the repetition rate of the pump laser. The ratio $e=p/B$ expresses the efficiency of each demultiplexer channel. Then for the described scheme $e = 0.225$, which does not include the efficiency of the detectors (0.85 on average), and photon losses due to the maximum possible frequency of the Pockels cell (4/6). With this in mind, the efficiency increases to $e=0.397$. This value is due to losses in the polarizing beam splitters, other optical elements of the circuit, the efficiency of optical couplers, and multiple passages through the Pockels cell.

\section{Discussion}

The characteristic feature of the demonstrated device is the optical circuit equipped with a single Pockels cell only. To the best of our knowledge, the majority of the demultiplexers reported to date include $N-1$ electro-optic active elements in order to split the input stream into $N$ channels (excluding probably the AOD-based design reported in \cite{Pont2022} which uses a single AOD). The number of channels which can be implemented using the demonstrated principle is limited by two parameters: the repetition rate $r$ of the source pump laser and the switching rate $r_s$ at which the Pockels cell or other electro-optic device can operate continuously. Typically the switching rate of the PC drivers is far below the repetition rate of pump lasers which can reach up to 10 GHz. The ratio $N_{min}\approx \lfloor r/r_s \rfloor$ defines the minimum number of channels of a loop-type demultiplexer. 
Although in our case the Pockels cell provides maximum switching rate $r_s$ of about 13~MHz (such type of cells are capable to produce $r_s$ well up to 20~MHz in burst mode) and in combination with $r=82.6$~MHz it gives $N_{min}=6$.
For this reason the use of slow PC drivers is impossible in this type of circuit because the geometry of an optical delay line required to store a large number of photons will quickly become impractical. In our case we can only fit 4 optical paths inside clear aperture of optical elements and hence we always lose 2 out of 6 photons. The demultiplexer can be easily upgraded to deliver a doubled amount of output streams if the repetition rate is also doubled. In this case each input pulse travels exactly half of the round-trip and when the PC is enabled two sets of beams pass the PC in different directions. One needs to add a PBS on the other side of PC to let both sets of streams out of the loop (see Appendix~\ref{app:demux-details}).

The loop-type demultiplexer benefits from self-synchronization of photons inside the loop. Once the PC rotates the polarization and releases the photons, they are already synchronous up to a few centimeter difference in the length of the output path to the fiber coupler. This means that there is no need in long delay lines which were used in the previous demonstrations \cite{Wang2019,Pont2022}. If the application demands the processing of the emitted photons with a photonic circuit then the input fibers of the fiber array can be directly connected to the fiber couplers thus eliminating any lossy fiber mating sleeves.

The disadvantage of the loop-type demultiplexer circuit is mostly related to the complexity of tuning the optical circuit and tailoring the selection of the elements in order to reach higher efficiency. We struggled with optimizing the position and orientation of the PC because two goals have to be achieved at once. The PC aperture must not cut the beams travelling inside the loop and the PC has to provide a high switching contrast at the same time. PCs with a larger clear aperture (our PCs have clear aperture equal to 3 mm) might simplify this task but this also leads to the increase of the half-wave voltage level which in turn makes it harder to reach fast switching rates. The second problem is the beam divergence due to intrinsic diffraction of the gaussian beam. The photons travel a large distance inside the loop and hence the coupling efficiency drops from the first channel to the fourth one. The solution would be to use a large focal length aspheric lens which deliver a larger collimated beam, however this again will require a larger PC aperture and hence a higher half-wave voltage. Nevertheless, we believe that careful optical engineering may significantly boost the demultiplexer efficiency. 



\section{Acknowledgements}
The work was supported by Russian Science Foundation grant 22-12-00353 (https://rscf.ru/en/project/22-12-00353/) in part of the demultiplexer design and characterization, and by Rosatom in the framework of the Roadmap for Quantum computing (Contract No. 868-1.3-15/15-2021 dated October 5, 2021 and Contract No.P2154 dated November 24, 2021). Quantum dot fabrication was supported by Rosatom in the framework of the Roadmap for Quantum computing (Contract No. 868-1.3-15/15-2021 dated October 5, 2021 and Contract No. R2152 dated November 19, 2021). S.K. and I.D. acknowledge support by the Interdisciplinary Scientific and Educational School of Moscow University "Photonic and Quantum Technologies. Digital Medicine".

\bibliography{bibliography}

\newpage

\appendix

\begin{figure}[htbp!]
	\includegraphics[width=0.99\linewidth]{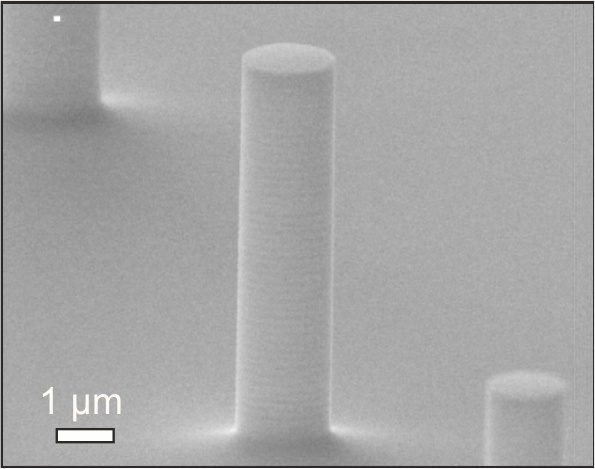}
	\caption{Scanning electron microscopy image of a typical micropillar.}
	\label{fig:SEM}
\end{figure}

\section{The QD source}\label{app:qd-source}

The quantum dot used in this work \cite{Rakhlin2022} was fabricated by molecular beam epitaxy on a GaAs:Si (001) substrate with a 500-nm GaAs buffer layer. Above the buffer layer, 28 (15) pairs of $\lambda$/4 Al$_{0.9}$Ga$_{0.1}$As/GaAs layers were grown and formed the lower (upper) distributed Bragg reflector. A 266-nm-thick GaAs $\lambda$ cavity was located between them; at the center of this cavity, an array of InAs QDs was formed with a surface density of 10$^9$-10$^{10}$ cm$^{-2}$ using the Stranski-Krastanov growth mechanism. The regular array of micropillars with a diameter of 1-3.5 $\mu$m was fabricated by reactive ion-plasma etching and standard contact lithography (365 nm) using a negative photoresist. A SEM image of the array is shown in Fig.~\ref{fig:SEM}. 

\begin{figure}[htbp!]
	\includegraphics[width=0.99\linewidth]{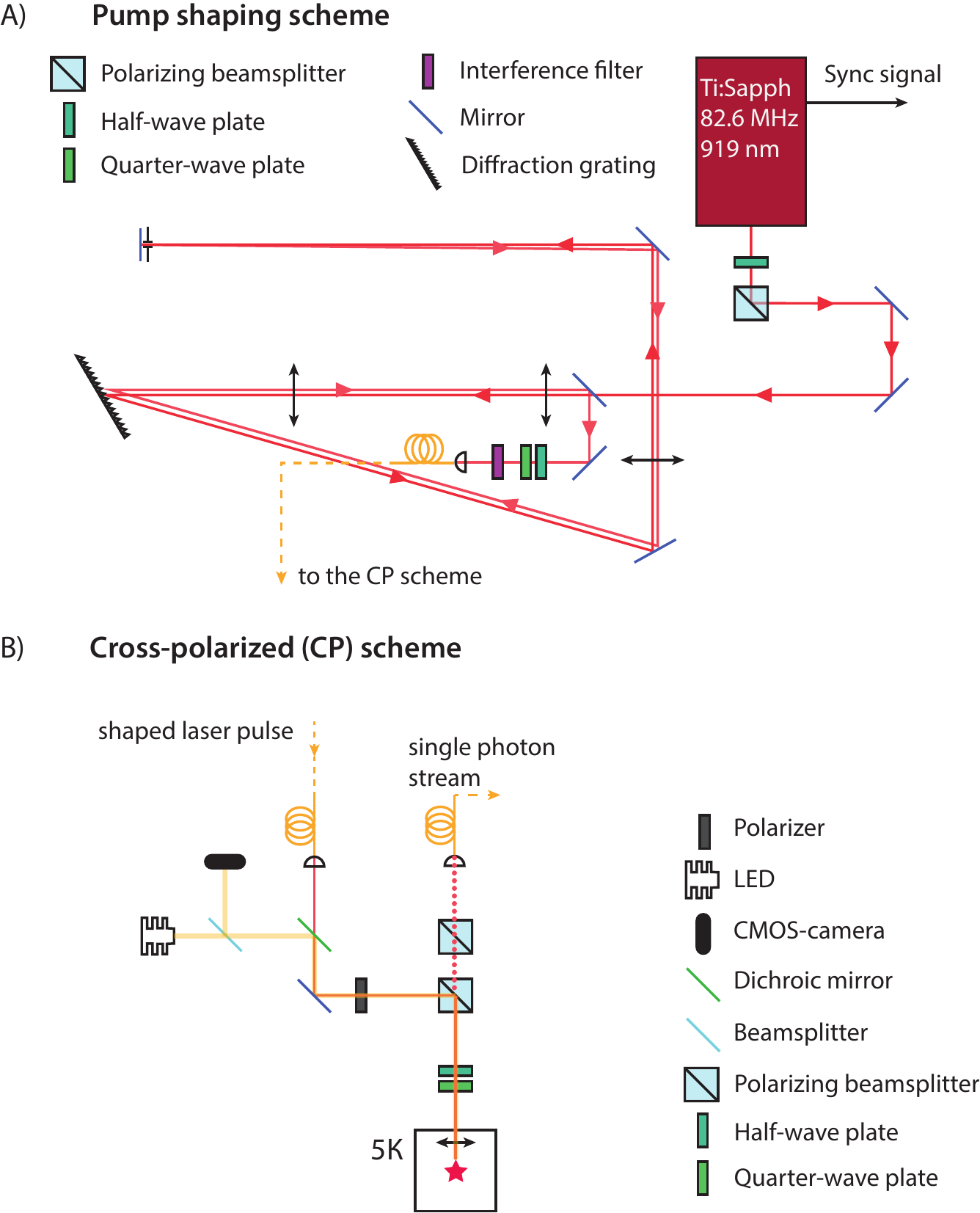}
	\caption{ a) Pulse shaping scheme. b) Cross-polarized scheme for pumping the quantum dot sample and rejecting pump light in the output channel.}
	\label{fig:single-photon-source-scheme}
\end{figure}

We used a resonant pumping regime to drive the quantum dot. Fig.\ref{fig:single-photon-source-scheme} a) illustrates the optical circuit used for spectral shaping of the pump pulse. We use a 4F scheme for spatial mapping of the spectral components of the initial femtosecond laser pulse. Then we use a variable slit to selectively reflect a portion of spectrum backwards. The backward propagating beam is slightly tilted in the vertical plane and we divert it towards a fiber coupler using a D-shaped mirror. The output shaped pulse is centered at 918.93 nm and has 0.10 nm width. The output single-mode fiber of the pulse-shaping circuit is then connected to the input of the cross-polarized scheme (see Fig.\ref{fig:single-photon-source-scheme} b)).

\section{Demultiplexer}\label{app:demux-details}

The optical memory cell is made in a form of a loop with a total length of 3.6~meters. The chosen length makes it possible to synchronize the photons entering the loop with those already inside the loop in time. The use of knife-edge prisms 

\begin{figure*}[htbp!]
	\includegraphics[width=2\columnwidth]{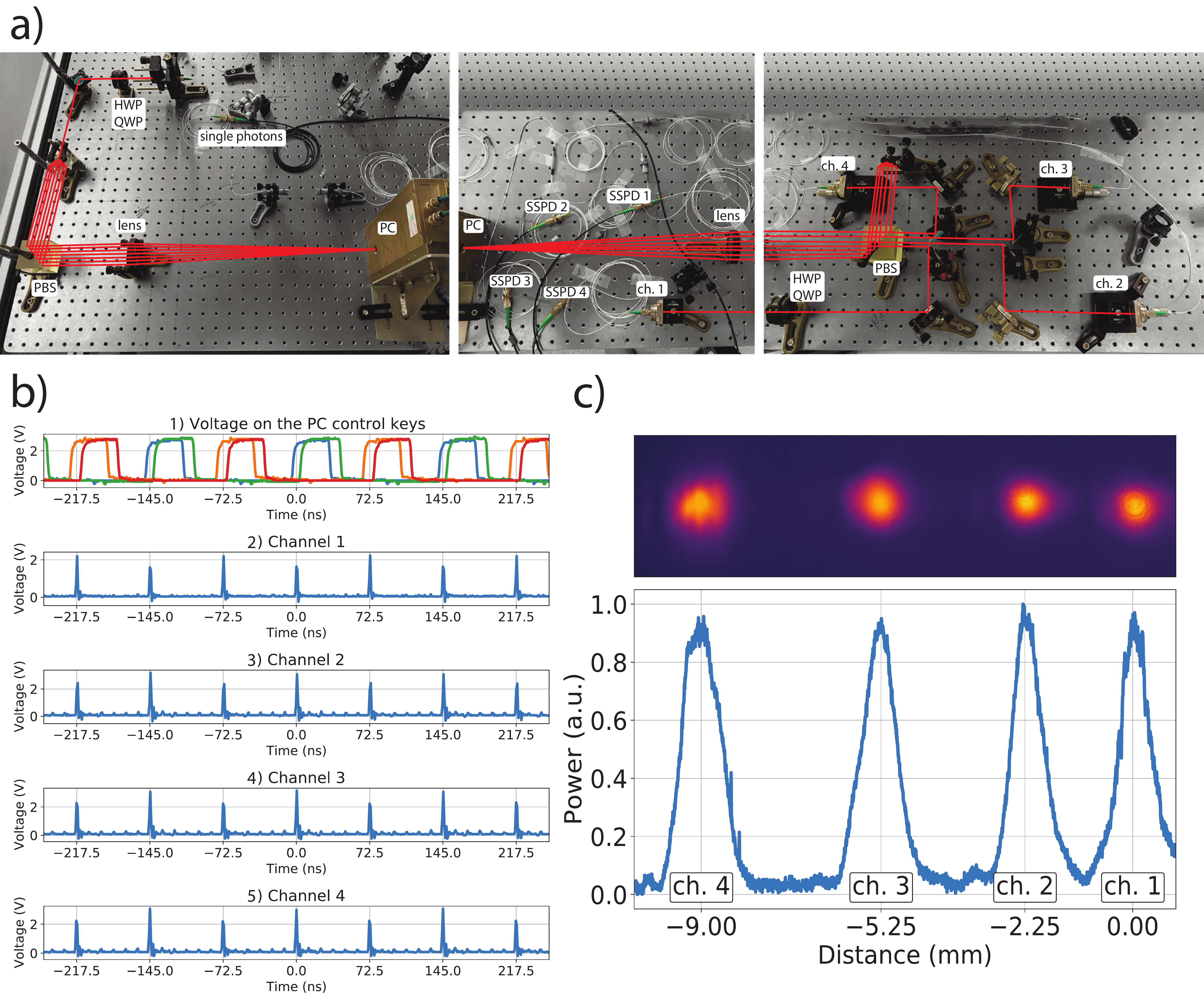}
	\caption{ a) Mode shapes measured at the output of the demultiplexer. An increase in the distance between optical modes occurs due to the non ideal mutual position of the reflecting prisms in the begining and the end of the loop. b) Principle of the PC driving. The figure illustrates the synchronization between electrical control pulses fed to the PC driver and the output  1) Voltage on control keys. 2-5) Electric pulses generated using analog detection of optical pules passing through a polarizing beamsplitter in each channel. c) Photo of the demultiplexer setup with beam paths marked. The whole setup did not fit into camera frame and had to be imaged in three shots. Red lines indicate approximate location of each optical path inside the circuit. The SSPD labels correspond to the connections between the output fiber-optic cables of the demultiplexor and the input cables of the SSPD detector.}
	\label{fig:setup_details}
\end{figure*}

makes it possible to separate the photons in the loop spatially, as shown in Fig.~\ref{fig:setup_details} a) and c) .

The Pockels cell (PC) is controlled by voltage pulses from an FPGA board, synchronized with the photon repetition rate. Figure~\ref{fig:setup_details} b.1) shows the sequence of voltage pulses sent to the control keys of the PC driver. The driver requires 4 TTL pulses to operate two high-voltage switches. The blue and orange curves indicate time moments when the first switch is turned on and off respectively, the green and red curves indicate time moments when the second switch is turned on and off respectively. In the time windows formed by the on/off switching of the keys, polarization of the photons rotates 90 degrees and they enter the loop. 

The outgoing photons are synchronized in time and separated in space, which allows them to be fed into optical couplers. An example of optical pulses in each channel measured with an analog photodetector is shown in Fig.~\ref{fig:setup_details} b) (panels 2-5). 

Since the photons travel a greater distance for each subsequent channel (3.6, 7.2, 10.8, and 14.4 meters, respectively), an increase in the diameter of the optical mode is observed as shown in Fig.~\ref{fig:setup_details} c). Also an increase in the distance between the optical modes is noticeable, which occurs due to the nonideal angles of the reflective prisms.

\section{Autocorrelation and HOM calculations}\label{app:auto-and-hom}

The autocorrelation function $g^{(2)}(0)$ is calculated by the formula
\begin{equation}
g^{(2)}(0) = \frac{A_0}{\langle A_i\rangle},
\end{equation}

where $A_0$ is the area of the central peak, $\langle A_i\rangle$ is the average area of the side peaks in the figure~\ref{fig:correlation}(a).

The uncorrected function of distinguishability of photons $HOM_{uncorr.}$ from different channels of the demultiplexer is given by the formula

\begin{equation}
HOM_{uncorr.} = 1 - \frac{2A_0}{\langle A_i\rangle},
\end{equation}
where $A_0$ and $\langle A_i\rangle$ are similar peak areas in figure~\ref{fig:HOM}.

The indistinguishability corrected for the features of the setup $HOM_{corr.}$ is given by the formula

\begin{equation}
HOM_{corr.} = \frac{1}{1-e^2} (1.5g^{(2)}(0) + \frac{R^2+T^2}{2RT} - \frac{R^2+T^2}{2RT}\frac{A_0}{\langle A_i\rangle}),
\end{equation}
where $e = 0.05$ is the visibility of the interference pattern measured for the classical light, $R=0.51$ and $T = 0.49$ are the reflection and transmission coefficients of the beamsplitter respectively, $A_0$ is the area of the central peak, $\langle A_i\rangle$ is the average area of the side peaks in the figure~\ref{fig:HOM}.

\end{document}